\documentclass[sigconf]{acmart}

\usepackage{xcolor}
\usepackage{booktabs}   
\usepackage{pifont}     
\usepackage{adjustbox}
\usepackage{url}
\usepackage{siunitx}
\usepackage{booktabs,multirow}
 
 \usepackage{tabularx}
\usepackage{graphicx}
\AtBeginDocument{%
  }

\setcopyright{acmlicensed}
\copyrightyear{2018}
\acmYear{2018}
\acmDOI{XXXXXXX.XXXXXXX}
\acmConference[Conference acronym 'XX]{Make sure to enter the correct
  conference title from your rights confirmation email}{June 03--05,
  2018}{Woodstock, NY}
\acmISBN{978-1-4503-XXXX-X/2018/06}

\begin{document}

\title{MMM-Fact: A Multimodal, Multi-Domain Fact-Checking Dataset with Multi-Level Retrieval Difficulty}


\author{Wenyan Xu}
\affiliation{%
  \institution{Central University of Finance and Economics}
  \city{Beijing}
  \country{China}
}
\email{2022211032@email.cufe.edu.cn}

\author{Dawei Xiang}
\affiliation{%
  \institution{University of Connecticut}
  \city{Storrs}
  \state{Connecticut}
  \country{USA}
}
\email{ieb24002@uconn.edu}

\author{Tianqi Ding}
\affiliation{%
  \institution{Baylor University}
  \city{Waco}
  \state{Texas}
  \country{USA}
}
\email{kirk\_ding1@baylor.edu}

\author{Weihai Lu}
\affiliation{%
  \institution{Peking University}
  \city{Beijing}
  \country{China}
}
\email{luweihai@pku.edu.cn}

\begin{abstract}
Misinformation and disinformation demand fact-checking that goes beyond simple evidence-based reasoning. Existing benchmarks fall short: they are largely single-modality (text-only), span short time horizons, use shallow evidence, cover domains unevenly, and often omit full articles—obscuring models’ real-world capability. We present \textbf{MMM-Fact} \footnote{\url{https://huggingface.co/datasets/Wenyan0110/MMM-Fact}}, a large-scale benchmark of 125{,}449 fact-checked statements (1995--2025) across multiple domains, each paired with the full fact-check article and multimodal evidence (text, images, videos, tables) from four fact-checking sites and one news outlet. To reflect verification effort, each statement is tagged with a retrieval-difficulty tier—Basic (1--5 sources), Intermediate (6--10), and Advanced ($>$10)—supporting fairness-amixedware evaluation for multi-step, cross-modal reasoning. The dataset adopts a three-class veracity scheme (true/false/not enough information) and enables tasks in veracity prediction, explainable fact-checking, complex evidence aggregation, and longitudinal analysis. Baselines with mainstream LLMs show MMM-Fact is markedly harder than prior resources, with performance degrading as evidence complexity rises. MMM-Fact offers a realistic, scalable benchmark for transparent, reliable, multimodal fact-checking. 
\end{abstract}



\maketitle

\section{Introduction}
\noindent \textit{Misinformation} and \textit{disinformation} cause substantial societal harm~\cite{sun2025audioenhancedvisionlanguagemodelinglatent, lu2025dammfnd}. The World Economic Forum’s \textit{Global Risks Report 2025}~\cite{elsner2025globalrisks} projects that “information disorder’’ will be the most severe global threat over the next two years. Fact-checking organizations respond by verifying dubious online statements and publishing evidence-based verdicts. A canonical workflow has three stages: (i) surfacing check-worthy claims, (ii) retrieving evidence, and (iii) evaluating claims against that evidence to produce a veracity judgment (e.g., ``true''/``false'') with an accompanying report~\cite{tong2024mmdfnd, altuncu2025factors}.

Despite progress, current pipelines strain under the internet’s volume and velocity~\cite{ENCODER, zhao2024balf}. Fact-checking is not a binary decision: it requires transparent sourcing and explicit reasoning, often aggregating multiple pieces of corroborating or refuting evidence across modalities (text, images, video, tables) and domains~\cite{PAIR, MEDIAN}. Policy frameworks echo these needs: the EU’s \textit{Digital Services Act}\footnote{\url{https://eur-lex.europa.eu/eli/reg/2022/2065/oj/eng}} and UNESCO’s \textit{Guidelines for Strengthening Trust in Media}\footnote{\url{https://www.unesco.org/en/internet-trust/guidelines}} emphasize multi-source verification and explainability. Accordingly, effective mitigation requires systems that perform multi-step reasoning over aggregated, multi-source evidence rather than one-shot retrieval\cite{ni2023feature,ni2025wonderfree}.

Evidence retrieval and reasoning difficulty also vary widely: some claims hinge on a single source; others require synthesizing dozens~\cite{xu2025finmultitime, xu2025learning}. Training or evaluating only on easy cases induces selection bias and inflates performance. Grading difficulty by required evidence (e.g., 1--5 vs.\ $\geq$10 pieces) better captures the spectrum from simple verification to complex, multi-step reasoning and enables fairer assessment~\cite{xiang2025promptsculptor, chen2025visrl}. As LLM capacity grows, large and diverse corpora are further needed to avoid overfitting and to improve robustness \cite{chen2025sifthinker,chen2025think, zhang2024yoloppa, zeng2025FSDrive}. 
\begin{figure}[!ht]
\centering
\includegraphics[width=1\linewidth]{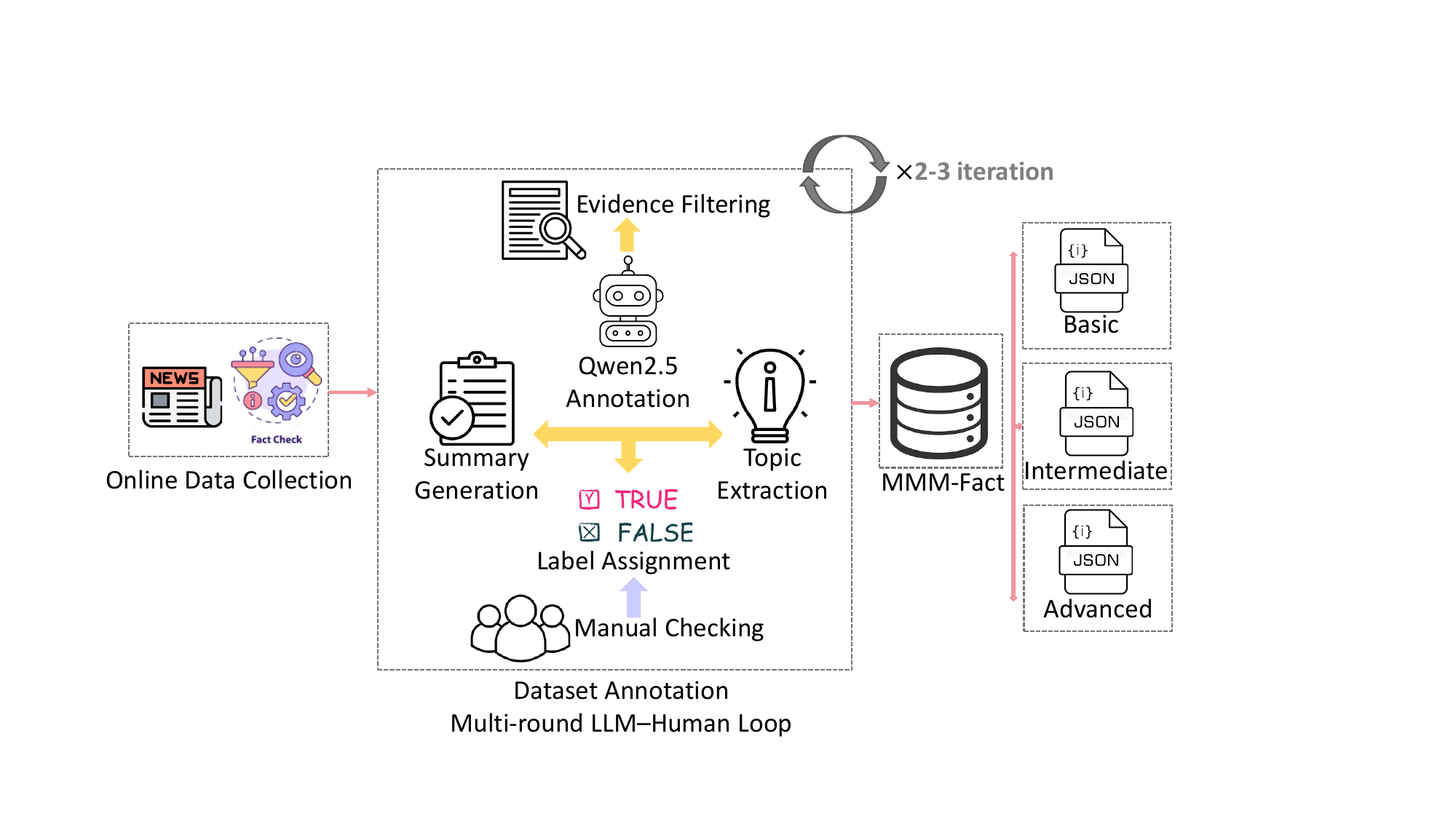}
\vspace{-7mm}
\caption{The MMM-Fact dataset contruction process.}
\label{fig:annotation}
\end{figure}

Existing datasets have advanced automated fact-checking, but most still exhibit \emph{limited modality coverage}. Many resources are derived from real-world claims yet remain predominantly text-centric: \textsc{PolitiFact}~\cite{vlachos2014fact}, \textsc{MultiFC}~\cite{augenstein2019multifc}, \textsc{AnswerFact}~\cite{zhang2020answerfact}, and \textsc{XFact}~\cite{gupta-srikumar-2021-x} focus on textual claims with text evidence or metadata; \textsc{FEVEROUS}~\cite{aly2021fact} adds tables; and \textsc{CHEF}~\cite{hu2022chef} and \textsc{MOCHEG}~\cite{yao2023end} extend to Chinese or cross-site sources. In practice, however, platforms mix text with charts, screenshots, and short videos. As a result, text-only benchmarks under-probe cross-modal alignment, image–text consistency, and visual provenance \cite{ni2025wonderturbo}. \textsc{FinFact}~\cite{rangapur2025fin} moves toward multimodality (text/image/metadata), but coverage remains incomplete. Most datasets also provide \emph{non-auditable, shallow evidence granularity}. Effective verification typically requires \emph{multi-step retrieval} and \emph{cross-source aggregation} across news, official databases, provenance checks, and third-party assessments, along with de-duplication and conflict resolution. Without an explicit notion of \emph{retrieval/aggregation difficulty}, evaluations skew toward “easy” cases and degrade on complex ones; even recent datasets such as \textsc{FACTors}~\cite{altuncu2025factors} and \textsc{ViFactCheck}~\cite{hoa2025vifactcheck} lack difficulty stratification, obscuring ceilings along the \emph{retrieve–rerank–aggregate–decide} pipeline.

Finally, many datasets operate in \emph{constrained domains or scale}. Several include claim-adjacent context but lack \emph{auditable evidence chains} and \emph{externally traceable links}, limiting interpretability and reproducibility (e.g., \textsc{FakeCovid}~\cite{shahi2020fakecovid}, \textsc{FakeNewsNet}~\cite{shu2020fakenewsnet}, \textsc{MuMiN}~\cite{nielsen2022mumin}, \textsc{MOCHEG}~\cite{yao2023end}, \textsc{ViFactCheck}~\cite{hoa2025vifactcheck}, \textsc{Podcasts}~\cite{setty2025annotation}).Many are also limited in size or temporal span, constraining cross-era robustness and longitudinal analyses. These gaps make models brittle under cross-modal, multi-hop, or contradictory evidence, and they hinder stability and reproducibility assessments over time. We highlight three application domains and situate our benchmark accordingly.
\begin{figure}[!ht]
\centering
\includegraphics[width=0.8\linewidth]{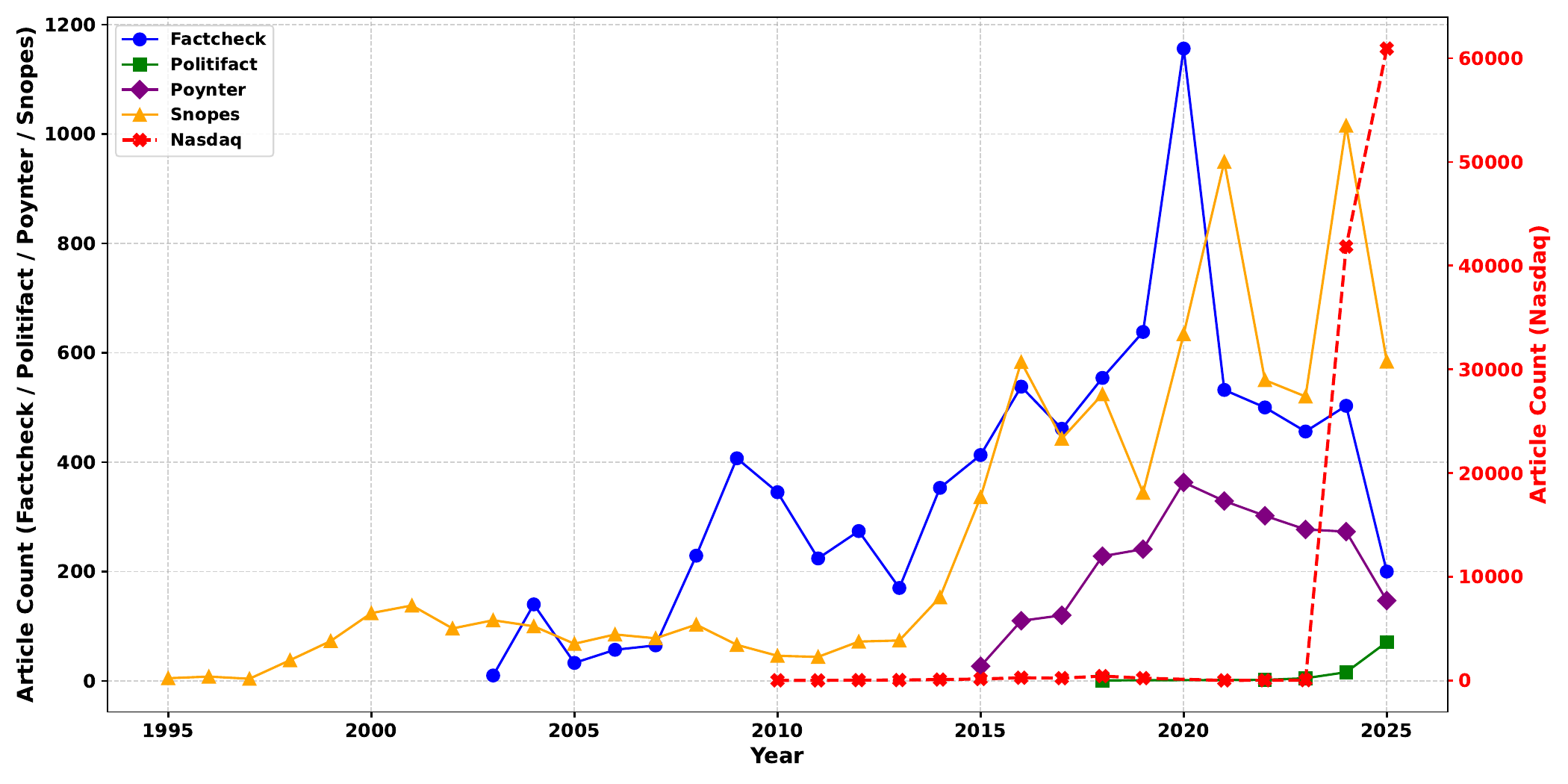}
\vspace{-4mm}
\caption{Yearly article counts for four fact-checking websites (Factcheck, Politifact, Poynter, and Snopes) and one news
website (Nasdaq).}
\label{fig:news_num}
\end{figure}

\begin{figure}[!ht]
\centering
\includegraphics[width=0.8\linewidth]{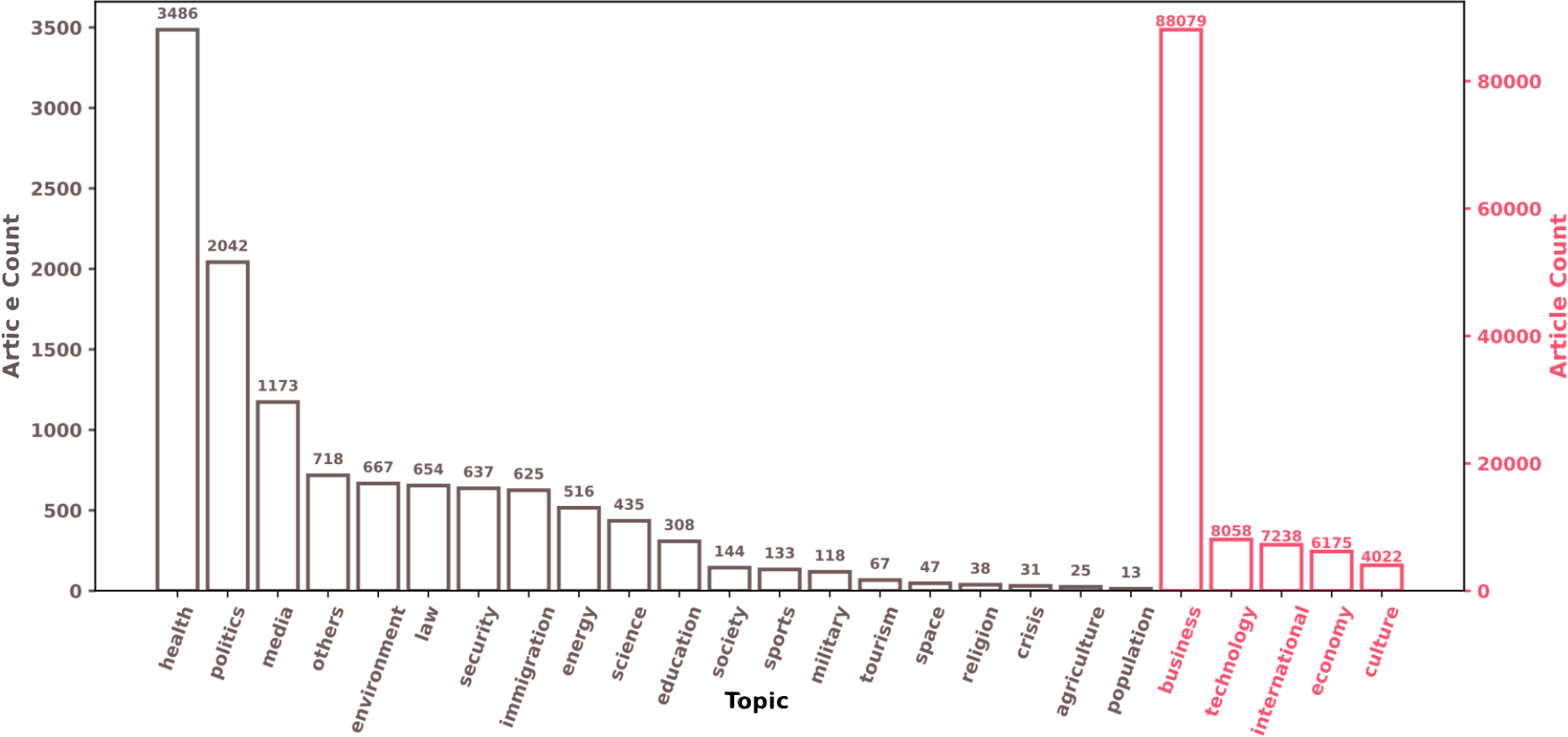}
\vspace{-5mm}
\caption{Distribution of Topics with Dual Y-Axes Highlighting Top Five Categories}
\label{fig:topic}
\end{figure}

In this paper, We introduce \textbf{MMM-Fact}, designed to close these gaps while aligning with real-world practice. MMM-Fact contains 125{,}449 statements fact-checked between 1995--2025, paired with complete fact-check articles. This 30-year scope enables longitudinal analyses across eras. The benchmark systematically incorporates multi-modal evidence---text, images, videos, and tables---and preserves \emph{auditable links} with \emph{paragraph-level localization}, supporting realistic end-to-end workflows (\emph{retrieve} $\rightarrow$ \emph{select} $\rightarrow$ \emph{cross-modal reasoning} $\rightarrow$ \emph{rationale}) as well as targeted module studies (e.g., OCR, reverse-image search, screenshot matching, table-fact extraction). Each claim is annotated with a difficulty tier: \textit{Basic} (1--5 evidence items, typically direct sources), \textit{Intermediate} (6--10, often requiring noise filtering), and \textit{Advanced} ($>10$ or highly diverse sources, often cross-source or multi-step). Finally, MMM-Fact adopts a three-class veracity scheme (``true,'' ``false,'' ``Not Enough Information (NEI)''), linking each statement to a full fact-check report detailing evidence and reasoning. Broad domain coverage (politics, health, economy, society, etc.) enables evaluation of concept drift, policy/office changes, and statistical updates, with era- and event-based splits for longitudinal study. Our contributions are:
\begin{enumerate}
\item We present MMM-Fact: 125{,}449 statements (1995--2025) spanning multiple domains, with complete fact-check articles and evidence for longitudinal and robustness studies.
\item We integrate text/image/video/table/metadata evidence from four fact-checking websites plus one news website, and introduce retrieval-difficulty labels (\textit{Basic} 1--5, \textit{Intermediate} 6--10, \textit{Advanced} $>$10) to enable cross-modal verification, multi-hop retrieval, and curriculum-style evaluation.
\item We provide baselines and systematic evaluations of mainstream LLMs on MMM-Fact, showing the benchmark’s difficulty and how performance degrades with increasing evidence complexity, thereby offering reproducible baselines and an analysis framework for future work.
\end{enumerate}

\section{The MMM-Fact Dataset}

To mitigate the gaps outlined above, we introduce \textsc{MMM-Fact}, a comprehensive benchmark for multimodal automated fact-checking and research on the full claim–context–evidence chain. The dataset contains 125{,}449 English fact-check instances drawn from five major sources—four fact-checkers (\textit{FactCheck}, \textit{PolitiFact}, \textit{Snopes}, \textit{Poynter}) and one news outlet (\textit{Nasdaq}). Each record includes standardized metadata (e.g., \texttt{Source\_Url}, \texttt{Claim}, \texttt{Author}, \texttt{Date}, \texttt{Summary}, \texttt{Article}, \texttt{Topic}, \texttt{Image}, \texttt{Evidence}, \texttt{Label}). Figure~\ref{fig:annotation} sketches the end-to-end pipeline (collection–cleaning–organization).

\subsection{Data Collection}

We built a reproducible, fault-tolerant crawler that honors \texttt{robots.txt} and rate limits, spanning October~19,~1995 to August~29,~2025.

\noindent\textbf{Snopes / FactCheck.org / PolitiFact / Nasdaq.} A unified two-stage pipeline first discovers articles (\emph{headline stage}) via keyword search and pagination, filtering URLs with a \texttt{/fact-check/} pattern and deduplicating. Headlines are extracted from \texttt{<h1>} with a slug fallback; results are serialized to JSONL for checkpointing. In the \emph{content stage}, stored URLs are revisited to extract body text, publication dates (from JSON-LD \texttt{datePublished}, normalized to ISO~8601), and images (from \texttt{og:image} and in-article \texttt{<img>}, preferring high-resolution absolute paths). A headless browser with strict rate control yields consistent UTF-8 JSONL.

\noindent\textbf{Poynter.} We use an API-first, HTML-fallback design. The headline stage queries the WordPress REST API (\texttt{/wp-json/wp/v2/posts}) with time-ordered pagination and deduplication, falling back to site scanning when necessary. The content stage prioritizes API text; otherwise, it parses \texttt{<article>} HTML (filtering newsletters/ \\ subscription blocks) and collects images from \texttt{data-src}/\texttt{srcset}. The pipeline is idempotent, auditable, and batch-executable.

Across sources, we initially collected 147{,}094 entries; after filtering and cleaning (\S\ref{sec:data_cleaning}), we consolidated 125{,}449 high-quality instances authored by 586 unique fact-checkers, with unified metadata and traceable evidence chains. We also distribute full article texts, not just metadata/URLs. MMM-Fact draws on publicly available content from five websites, crawled in accordance with each site’s robots.txt and usage terms.

\subsection{Data Cleaning and Preparation}
\label{sec:data_cleaning}

Cleaning proceeds in reproducible stages (Figure~\ref{fig:annotation}), assisted by \texttt{Qwen2.5-7B-Instruct} with a 15\% random manual spot-check.

\begin{itemize}
    \item \textbf{Field \& length checks:} Drop items missing title/body/claim/ \\ verdict; remove claims or bodies $<40$ chars.
    \item \textbf{Date normalization:} Convert all times to "YYYY-MM-DD".
    \item \textbf{Topic assignment:} Case-insensitive classification over an extended lexicon; select the top label across 25 categories (Figure~\ref{fig:topic}).
    \item \textbf{Two-sentence summaries:} Deterministic prompts yield exactly: ``\emph{Claim to verify: ...}'' and ``\emph{Rationale: ... (Verdict: ...)},'' followed by year/punctuation/whitespace normalization.
    \item \textbf{Evidence extraction:} Parse \texttt{<article>/<main>/div.arti \\ cle\_\_content}; segment sentences; map each hyperlink to its sentence; merge sentences with identical link sets into evidence units \{\texttt{sentence}, \texttt{hrefs[]}\}; normalize URLs; filter promotional/irrelevant content.
    \item \textbf{Difficulty tags:} Remove empty evidence; label by evidence count —\textit{basic} (1–5), \textit{mid-level} (6–10), \textit{advanced} ($>$10).
    \item \textbf{Normalization \& deduplication:} Strip HTML/emoji/escapes; normalize whitespace; remove duplicate paragraphs.
    \item \textbf{Label standardization:} Map heterogeneous ratings (e.g., \emph{true}, \emph{false}, \emph{satire}, \emph{misleading}, \emph{unknown}) to \{True, False, Not Enough Information (NEI)\}; case- and phrase-aware matching (e.g., ``This claim is true.'' $\rightarrow$ True); unmatched $\rightarrow$ NEI.
\end{itemize}

\subsection{Dataset Statistics}

Core fields (\texttt{claim\_title}, \texttt{analysis}, \texttt{rating}) show near-complete coverage. The \textit{Nasdaq} and \textit{FactCheck} slices contribute the bulk of the records; \textit{Snopes} ranks among the top few sources by record count. Evidence domains are diverse: finance/media sites (e.g., \textit{barchart.com}, \textit{nasdaq.com}, \textit{fool.com}) dominate, while \textit{factcheck.org}, \textit{snopes.com}, \textit{politifact.com}, and government sources account for a substantial share, yielding a balanced mix of news, finance, and verification outlets. The collection includes text links, with video links predominating. Evidence difficulty varies widely: \textit{basic} accounts for 73{,}477 cases (58.57\%), \textit{mid-level} for 21{,}873 (17.44\%), and \textit{advanced} ($>$10 links) for 30{,}099 (23.99\%), underscoring substantial heterogeneity in citation density.
Overall, \textsc{MMM-Fact} pairs scale with diversity and reasoning complexity, offering a unified, auditable benchmark for multimodal, verifiable fact-checking. 

\begin{table}[t]
\centering
\caption{Model performance (Precision, Recall, F1) across difficulty levels (Basic, Mid-level, and Advanced). Bold values indicate the best scores within each column.}
\vspace{-4mm}
\label{tab:model-perf}
\small
\setlength{\tabcolsep}{3pt} 
\scalebox{0.8}{ 
\begin{tabular}{l l ccc ccc ccc}
\toprule
& & \multicolumn{3}{c}{\textbf{Basic}} & \multicolumn{3}{c}{\textbf{Mid-level}} & \multicolumn{3}{c}{\textbf{Advanced}} \\
\textbf{Family} & \textbf{Model} & \textbf{Prec.} & \textbf{Rec.} & \textbf{F1} & \textbf{Prec.} & \textbf{Rec.} & \textbf{F1} & \textbf{Prec.} & \textbf{Rec.} & \textbf{F1} \\
\midrule
\multirow{4}{*}{\textbf{NLI (Text)}}
 & \textbf{ALBERT}      & 0.495 & 0.397 & 0.441 & 0.431 & 0.338 & 0.379 & 0.396 & 0.327 & 0.358 \\
 & \textbf{RoBERTa-L}   & 0.442 & 0.387 & 0.413 & 0.353 & 0.270 & 0.306 & 0.359 & 0.334 & 0.346 \\
 & \textbf{BART-L}      & 0.402 & 0.362 & 0.381 & 0.378 & 0.346 & 0.361 & 0.375 & 0.353 & 0.364 \\
 & \textbf{ELECTRA}     & 0.328 & 0.379 & 0.352 & 0.353 & 0.359 & 0.356 & 0.401 & 0.350 & 0.374 \\
\midrule
\multirow{4}{*}{\textbf{LLM (Text)}}
& \textbf{GPT-4} & \textbf{0.775} & \textbf{0.776} & \textbf{0.776} & \textbf{0.702} & \textbf{0.697} & \textbf{0.699} & \textbf{0.658} & \textbf{0.734}
& \textbf{0.694} \\
& \textbf{LLaVA}        & 0.722 & 0.703 & 0.712 & 0.590 & 0.618 & 0.604 & 0.400 & 0.403 & 0.401 \\
& \textbf{DeepSeek}  & 0.717 & 0.697 & 0.707 & 0.550 & 0.620 & 0.583 & 0.413 & 0.429 & 0.421 \\
& \textbf{Doubao} & 0.605 & 0.597 & 0.601 & 0.448 & 0.323 & 0.376 & 0.428 & 0.438 & 0.433 \\
\bottomrule
\end{tabular}
}
\end{table}

\begin{table}[t]
\centering
\caption{F1 scores by model and prompting strategy across difficulty levels and evidence modalities (higher is better). "—" indicates a configuration not evaluated.}
\vspace{-4mm}
\label{tab:f1-by-modality}
\small
\setlength{\tabcolsep}{2pt}
\scalebox{0.8}{
\begin{tabular}{l l cc cc cc}
\toprule
& & \multicolumn{2}{c}{\textbf{Basic}} & \multicolumn{2}{c}{\textbf{Mid-level}} & \multicolumn{2}{c}{\textbf{Advanced}} \\
\textbf{Model} & \textbf{Strategy} & \textbf{Text \& Image} & \textbf{Text} & \textbf{Text \& Image} & \textbf{Text} & \textbf{Text \& Image} & \textbf{Text} \\
\midrule
\multirow{3}{*}{\textbf{LLaVA}}
  & CoT        & 0.700 & 0.586 & \textbf{0.741} & 0.565 & \textbf{0.673} & 0.487 \\
  & Symbolic   & 0.499 & 0.404 & 0.498 & 0.417 & 0.445 & 0.402 \\
  & Self-Help  & 0.297 & 0.299 & 0.277 & 0.344 & 0.198 & 0.365 \\
\midrule
\multirow{3}{*}{\textbf{GPT-4}}
  & CoT        & 0.779 & 0.606 & 0.576 & 0.570 & 0.519 & 0.489 \\
  & Symbolic   & \textbf{0.805} & \textbf{0.612} & 0.579 & 0.141 & 0.507 & 0.516 \\
  & Self-Help  & 0.762 & \textbf{0.612} & 0.594 & 0.582 & 0.494 & 0.540 \\
\midrule
\multirow{3}{*}{\textbf{Qwen}}
  & CoT        & 0.577 & —     & 0.632 & —     & 0.655 & —     \\
  & Symbolic   & 0.491 & —     & 0.539 & —     & 0.547 & —     \\
  & Self-Help  & 0.365 & —     & 0.344 & —     & 0.416 & —     \\
\midrule
\multirow{3}{*}{\textbf{DeepSeek}}
  & CoT        & —     & 0.583 & —     & 0.576 & —     & 0.561 \\
  & Symbolic   & —     & 0.559 & —     & 0.555 & —     & 0.547 \\
  & Self-Help  & —     & 0.468 & —     & 0.456 & —     & 0.422 \\
\midrule
\multirow{3}{*}{\textbf{Doubao}}
  & CoT        & —     & 0.595 & —     & \textbf{0.589} & —     & 0.486 \\
  & Symbolic   & —     & 0.585 & —     & 0.580 & —     & \textbf{0.577} \\
  & Self-Help  & —     & 0.486 & —     & 0.506 & —     & 0.475 \\
\bottomrule
\end{tabular}
}
\end{table}

\section{Evaluation and Analysis}
\label{sec:evaluation}

\subsection{Performance Across Difficulty Levels}

Motivated by rapid advances in large language models, we run direct inference on the MMM-Fact evaluation benchmark with vision–language models (e.g., GPT-4V, LLaVA), text-only LLMs (e.g., DeepSeek, Doubao), and NLI baselines; therefore, we do not provide official train/dev/test splits.

Table~\ref{tab:model-perf} reports Precision, Recall, and F1 across three difficulty levels.  
Among text-only NLI baselines, \textbf{ALBERT} attains the highest Basic F1 (0.441), while \textbf{ELECTRA} slightly leads in the Advanced tier (0.374). All show a consistent decline in recall and F1 as difficulty rises, reflecting limited ability for multi-step or context-rich reasoning. Large multimodal LLMs display stronger robustness. \textbf{GPT-4} delivers the best overall performance, far surpassing other models. The moderate drop reflects the growing reasoning demands of longer, more complex evidence chains rather than overfitting to simpler inputs.  
\textbf{LLaVA} and \textbf{DeepSeek} remain competitive at mid-level but degrade in Advanced tasks, indicating challenges in integrating heterogeneous evidence. \textbf{Doubao} shows moderate stability yet lower recall, suggesting less effective evidence aggregation.

\subsection{Impact of Prompting Strategy and Modality}

Table~\ref{tab:f1-by-modality} compares prompting strategies (CoT, Symbolic, Self-Help) and modalities (Text vs. Text \& Image).  
CoT consistently yields the strongest results for \textbf{LLaVA} and \textbf{GPT-4}, confirming that explicit reasoning steps improve factual grounding.  
Symbolic reasoning benefits \textbf{GPT-4}, achieving the top Basic F1 (0.805) and stable Advanced performance, indicating better structure-aware generalization.  
Self-Help performs weakest across models, showing that unguided reasoning often leads to hallucinations and incomplete retrieval.  
Across all systems, Text \& Image inputs outperform Text-only settings, particularly in harder tiers, underscoring the role of cross-modal alignment in complex claim verification. Overall, results reveal that (1) multimodal LLMs substantially outperform text-only NLI models, (2) reasoning-guided prompting—especially CoT and Symbolic—is critical for multi-hop inference, and (3) performance consistently declines with evidence complexity, highlighting ongoing challenges in long-context, cross-modal reasoning.
 
\section{Conclusion}

\textbf{MMM-Fact} is a large-scale benchmark that addresses persistent gaps in prior work—single-modality evidence, short time spans, shallow evidence, uneven domain coverage, and missing full articles. Spanning 1995–2025, it links 125{,}449 real-world claims to full fact-check articles and \textit{multimodal} evidence (text, images, video, tables). It also annotates \textit{retrieval difficulty} (Basic/Intermediate/Advanced) and uses a three-class veracity scheme aligned with professional practice, enabling fairness-aware evaluation and curriculum-style training for multi-hop, cross-modal reasoning. Baselines with mainstream LLMs show MMM-Fact is substantially harder than prior datasets, with performance declining as evidence complexity rises. These results establish MMM-Fact as a rigorous testbed for \textit{explainable fact-checking}, \textit{multi-step retrieval}, \textit{cross-modal reasoning}, and \textit{longitudinal} analysis.

\bibliographystyle{ACM-Reference-Format}
\bibliography{sample-base}




\end{document}